\documentclass[prl,twocolumn]{revtex4}

\usepackage{amsmath}
\usepackage{graphicx}
\usepackage{epsfig}
\usepackage{epstopdf}

\begin{document}
\title{Azimuthal correlations of forward di-hadrons in d+Au collisions at RHIC\\in the Color Glass Condensate}

\author{J.L. Albacete}
\email{javier.lopez-albacete@cea.fr}
\author{C. Marquet}
\email{cyrille.marquet@cea.fr}
\affiliation{Institut de Physique Th{\'e}orique, CEA/Saclay, 91191 Gif-sur-Yvette cedex, France\\
URA 2306, unit\'e de recherche associ\'ee au CNRS}

\begin{abstract}
We present a good description of recent experimental data on forward di-hadron azimuthal correlations measured in deuteron-gold collisions at RHIC, where {\it monojet} production has been observed. Our approach is based on the Color Glass Condensate  theory for the small-$x$ degrees of freedom of the nuclear wave function, including the use of non-linear evolution equations with running QCD coupling. Our analysis provides further evidence for the presence of saturation effects in RHIC data.
\end{abstract}

\maketitle

Analyses of the angular dependence of two-particle correlations in hadronic collisions have proven
to be an essential tool for testing the underlying QCD microscopic dynamics. Indeed, the presence of well pronounced back-to-back azimuthal correlations for high transverse momentum particles in proton-proton (p+p) collisions at RHIC and Tevatron \cite{Abazov:2004hm} is well understood within the standard leading-twist approach to QCD. There, they originate from the scattering of two free partons, each collinear to one of the colliding protons, together with energy-momentum conservation. In turn, the absence of back-to-back correlations observed in gold-gold collisions at RHIC \cite{Adams:2003im}, for mid-rapidity pairs, has been interpreted as due to the interaction of the outgoing partons with a dense and strongly interacting medium produced in such collisions, presumably a Quark Gluon Plasma.

Related measurements have also been performed in deuteron-gold (d+Au) collisions at RHIC. While for mid-rapidity pairs data is consistent with a back-to-back structure, recent measurements of di-hadron correlations in the deuteron fragmentation region, i.e. at forward rapidities, display a strong suppression of the away-side peak \cite{Braidot:2010zh}, a phenomenon also referred to as {\it monojet} production. The presence of back-to-back correlations at mid-rapidity and the fact that total particle multiplicities decrease at forward rapidities rule out a {\it final state} interpretation of this novel phenomenon. Rather, they suggest an {\it initial state} effect or a modification in the primary particle production mechanism with respect to the standard collinear factorization approach.

At high energies, particle production in the forward rapidity region probe partons carrying a small fraction $x$ of the target (nucleus) longitudinal momentum. Moreover, the gluon densities within a hadron grow very fast with decreasing $x$. Therefore, and contrary to the assumptions of collinear factorization, gluons in such regime cannot be regarded as independent. Rather, they form a highly coherent system governed by non-linear phenomena and gluon saturation. These qualitative ideas are casted in a precise theoretical framework known as the Color Glass Condensate (CGC) effective theory of QCD in the high-energy limit (see e.g. \cite{Gelis:2010nm} and references therein).

There is compelling evidence that gold-gold and d+Au collisions at RHIC probe the saturation
regime of QCD. This claim is supported by the success of saturation models \cite{Albacete:2007sm,Kharzeev:2002pc} in the description
of the energy, rapidity and centrality dependence of the transverse-momentum integrated particle multiplicities experimentally measured. On the other hand, CGC-based calculations successfully predicted first \cite{Kharzeev:2003wz,Albacete:2003iq}, and correctly described then \cite{Dumitru:2005gt,Albacete:2010bs}, the measured suppressed production of high transverse-momentum forward hadrons in d+Au collisions with respect to p+p collisions \cite{Arsene:2004ux,Adams:2006uz}. Despite the success of CGC phenomenology at RHIC outlined above, alternative interpretations of data are also possible \cite{alternatives}. Although some of these approaches invoke dynamical ingredients related to those in the CGC, such as unitarity or multiple scatterings, it remains unsettled whether the latter is indeed the right approximation to QCD that better grasps the relevant dynamics probed at RHIC.

Disentangling the presently unclear situation requires the study of more exclusive observables. Forward di-hadron correlations in d+Au collisions are a good testing ground for models. Indeed, the possibility of {\it monojet} production as a result of saturation effects was anticipated in \cite{Kharzeev:2004bw,Marquet:2007vb}, albeit at a qualitative level. In this letter we shall present a CGC analysis, using the most up-to-date theoretical tools available, of the recent data on forward di-hadron azimuthal correlations in d+Au collisions at RHIC by the STAR collaboration, providing a good description of the data. We argue that our results provide further evidence for the presence of saturation effects in RHIC data and also that the CGC effective theory approximates well QCD in the saturation regime.

We are interested in the calculation of the double-inclusive hadron production cross section in proton-nucleus (p+A) collisions $d\sigma^{p\,A\to h_1 h_2 X}$. The theoretical groundwork needed for this calculation in the CGC was presented in \cite{JalilianMarian:2004da,Nikolaev:2005dd,Baier:2005dv,Marquet:2007vb}. We shall denote $p_{1,2\,\perp}$ and $y_{1,2}=\ln(p_{1,2}^+/p_{1,2}^-)/2$ the transverse momenta and rapidities of the produced hadrons, using light-cone coordinates for four-dimensional vectors $p=(p^+,p^-,p_{\perp})$. We introduce the Feynman-$x$ variables $x_i=|p_{i\perp}|e^{y_i}/\sqrt{s},$ with $\sqrt{s}$ the collision energy per nucleon. The kinematics of di-hadron production is such that the partons in the proton (nucleus) wave function that can contribute to the cross section carry a fraction of longitudinal momentum bounded from below by $x_p$ ($x_A$), given by
\begin{equation}
x_p=x_1+x_2\ ,\hspace{0.5cm}
x_A=x_1\ e^{-2y_1}+x_2\ e^{-2y_2}\ .
\label{kin}
\end{equation}
The STAR measurements to be analyzed here feature $\sqrt{s}=200$ GeV $\gg |p_{i\perp}|$ and $y_{i}\sim 3$, yielding $x_p\!\sim\!0.4$ and $x_A\!\sim\!10^{-3}$. Thus, the scattering process involves quantum fluctuations well understood in QCD on the proton side, while, on the nuclear side it probes quantum fluctuations whose non-linear QCD dynamics can be studied.

First measurements of the coincidence probability were performed at more central rapidities \cite{Adams:2006uz,Adler:2006hi}, less suited for exploration of small-$x$ effects. These purely kinematic considerations motivated the approach of \cite{Marquet:2007vb}, which we shall follow here. There, the proton is treated as a collinear object and described by means of the usual parton distribution functions, whereas the nucleus target is probed in the high parton density regime and is amenable to a CGC description. In this picture, the scattering process is initiated by a fast valence quark from the incoming proton (or deuteron) which splits into a quark and a gluon, either before or after scattering off the saturated small-$x$ glue of the target. Importantly, the splitting into a quark-gluon pair is described by standard pQCD techniques and, therefore, features an exact back-to-back correlation in the transverse plane. The quark-gluon system is put on-shell through the interaction with the nucleus, as a result of which the quark and gluon also acquire a transverse momentum of the order of the saturation scale of the nucleus. When that scale, which marks the onset of non-linear effects, is comparable to the initial transverse momenta of the quark and gluon, their intrinsic angular correlation is washed out. Finally, the outgoing quark and gluon fragment independently into hadrons.

The forward di-hadron production yield in p+A collisions reads
\begin{eqnarray}
dN^{pA\to h_1 h_2 X}=\int_{x_1}^1 dz_1 \int_{x_2}^1 dz_2 \int_{\frac{x_1}{z_1}+\frac{x_2}{z_2}}^1 dx\ f_{q/p}(x,\mu)\nonumber\\ \left[dN^{qA\to qgX}\left(xP,\frac{p_1}{z_1},\frac{p_2}{z_2}\right)D_{h_1/q}(z_1,\mu)D_{h_2/g}(z_2,\mu)+\right.\nonumber\\\left.
dN^{qA\to qgX}\left(xP,\frac{p_2}{z_2},\frac{p_1}{z_1}\right)D_{h_1/g}(z_1,\mu)D_{h_2/q}(z_2,\mu)\right]\ ,
\label{collfact}
\end{eqnarray}
where $f_{q/p}$ and $D_{h/i}$ represent the valence quark distribution of the proton and the fragmentation function of a parton $i$ into hadron $h$ respectively. 
We use the CTEQ6 NLO quark distributions \cite{Pumplin:2002vw} and the KKP NLO fragmentation functions \cite{Kniehl:2000fe}. In the analysis of d+Au collisions we shall neglect all nuclear effects in the deuteron, regarded as an incoherent superposition of a proton and a neutron. We obtain the neutron valence quark distribution from the proton one using isospin symmetry.
The factorization and fragmentation scales are both chosen equal to the transverse momentum of the leading hadron, which we denote as hadron 1, $\mu=|p_{1\perp}|.$
Finally, the cross section for the inclusive production of the quark-gluon system in the scattering of a quark with momentum $xP^+$ off the nucleus $A$ reads \cite{Marquet:2007vb}:
\begin{eqnarray}
\frac{dN^{qA\to qgX}}{d^3kd^3q}=
\frac{\alpha_S C_F}{4\pi^2}\ \delta(xP^+\!-\!k^+\!-\!q^+)\ F(\tilde{x}_A,\Delta)
\nonumber\\\sum_{\lambda\alpha\beta}
\left|I^{\lambda}_{\alpha\beta}(z,k_\perp\!-\!\Delta;{\tilde{x}_A})\!-\!
\psi^{\lambda}_{\alpha\beta}(z,k_\perp\!-\!z\Delta)\right|^2\ ,
\label{cs}\end{eqnarray}
where $P$, $q$ and $k$ denote the deuteron, quark and gluon momenta, respectively. We denoted $\Delta=k_\perp+q_\perp$, $z=k^+/xP^+$ and the longitudinal momentum fraction of gluons in the nucleus $\tilde{x}_A=x_1\ e^{-2y_1}/z_1+x_2\ e^{-2y_2}/z_2$. Due to parton fragmentation the values of $x$ probed are generically higher than $x_p$ and $x_A$ defined in Eq.~\eqref{kin}. For the proton, one has $0.4\sim x_p<x<1$. In that region the proton wave function is dominated by valence quarks. Therefore we have safely neglected the gluon initiated processes $gA\to q\bar{q}X$ and $gA\to ggX$ in Eq.~\eqref{collfact}.

Let us review the ingredients in Eq.~\eqref{cs}. $F(x,k_\perp)$ is the unintegrated gluon distribution (ugd) of the target nucleus. It is related to the correlator $\mathcal{N}$ of two light-like fundamental Wilson lines in the background of the color fields of the nucleus, also referred to as dipole scattering amplitude, through a Fourier transform:
\begin{equation}
F(x,k_\perp)=\int\frac{d^2r}{(2\pi)^2}\ e^{-ik_\perp\cdot r}\ [1-{\cal N}(x,r)]\ ,
\label{ugd}
\end{equation}
where $r$ denotes the dipole transverse size. The second line of Eq.~\eqref{cs} features the $k_T$-factorization breaking term. Its exact evaluation requires the knowledge of four and six-point correlators of Wilson lines \cite{JalilianMarian:2004da,Nikolaev:2005dd,Baier:2005dv,Marquet:2007vb}, which are difficult to obtain in practice. However, assuming that the distribution of color charges in the nucleus is Gaussian with a non-local variance, higher-point correlators can be expressed in terms of the two-point function $F(x,k_\perp)$. As shown in \cite{Marquet:2007vb}, further use of the large-$N_c$ limit yields the following expression for $I$:
\begin{equation}
I^{\lambda}_{\alpha\beta}(z,k_\perp;x)
=\int d^2q_\perp \psi^{\lambda}_{\alpha\beta}(z,q_\perp) F(x,k_\perp\!-\!q_\perp)\ ,
\label{split}\end{equation}
where $\psi^{\lambda}_{\alpha\beta}$ is the amplitude for $q\!\to\!qg$ splitting and $\lambda,$ $\alpha$ and $\beta$ are polarization and helicity indices. We remark that, even under the Gaussian approximation, the cross section \eqref{cs} is still a non-linear function of the unintegrated gluon distribution, thereby invalidating $k_T$-factorization, namely $dN^{qA\to qgX}\neq F\otimes dN^{qg\to qgX}$.
Finally, the delta function in Eq.~\eqref{cs} is due to the use of the eikonal approximation. It reflects the fact that in a high-energy scattering, the momentum transfer is mainly transverse, since the exchange of longitudinal momentum is suppressed by powers of the collision energy.

The CGC is endowed with a set of non-linear evolution equations which in the large-$N_c$ limit reduce to the Balitsky-Kovchegov (BK) equation \cite{Balitsky:1995ub,Kovchegov:1999yj}. These equations are renormalization group equations for the $x$ evolution of the unintegrated gluon distribution, and more generally of $n$-point correlators, in which both linear radiative processes and non-linear {\it recombination} effects. In this work, we compute the small-$x$ dynamics of the dipole correlator, and hence that of the ugd, by solving the BK equation including running coupling corrections (rcBK):
\begin{eqnarray}
  \frac{\partial {\cal N}(x,r)}{\partial\ln(x_0/x)}=\int d^2r_1\
  K^{{\rm run}}(r,r_1,r_2) \left[{\cal N}(x,r_1)\right.\nonumber\\\left.
+{\cal N}(x,r_2)-{\cal N}(x,r)- {\cal N}(x,r_1)\,{\cal N}(x,r_2)\right]\ .
\label{bk1}
\end{eqnarray}
The evolution kernel $K^{{\rm run}}$ is evaluated according to the prescription of \cite{Balitsky:2006wa}. Explicit expressions for the kernel, together with a detailed discussion on the numerical method used to solve the rcBK equation can be found in \cite{rcBK}. The only piece of information left to fully complete all the ingredients in Eq.~\eqref{collfact} are the initial conditions for the rcBK evolution. This non-perturbative input has been constrained by the analysis of single-inclusive forward hadron production in d+Au collisions at RHIC performed in  \cite{Albacete:2010bs} using an analogous CGC set up: the starting point of the rcBK evolution is $x_0=0.02$, and at that value of $x$ the initial saturation scale probed by quarks is $\bar{Q}_{s0}^2=0.4$ GeV$^2$. Here, we simply take over this information. In this respect, the forward di-hadron calculation presented here is parameter-free.

We will now investigate the process $dAu\!\to\!h_1h_2X,$ with $\sqrt{s}\!=\!200\ \mbox{GeV}.$ More specifically, we are interested in the coincidence probability, an experimental quantity measured by both the PHENIX and STAR collaborations at RHIC. It is given by $CP(\Delta\phi)=N_{pair}(\Delta\phi)/N_{trig}$ with
\begin{equation}
N_{pair}(\Delta\phi)=\hspace{-0.4cm}\int\limits_{y_i,|p_{i\perp}|}\hspace{-0.3cm}
\frac{dN^{dAu\to h_1 h_2 X}}{d^3p_1 d^3p_2}\ ,\
N_{trig}=\hspace{-0.3cm}\int\limits_{y,\ p_\perp}\hspace{-0.2cm}\frac{dN^{dAu\to hX}}{d^3p}\ ,
\label{cp}
\end{equation}
and it has the meaning of the probability of, given a trigger hadron $h_1$ in a certain momentum range, produce an associated hadron $h_2$ in another momentum range and with a difference between the azimuthal angles of the two particles equal to $\Delta\phi$. In order to study the centrality dependence of the coincidence probability, we identify the centrality averaged initial saturation scale $\bar{Q}^2_{s0}$, extracted from minimum-bias single-inclusive hadron production data, with the value of $Q^2_{s0}$ at $b=5.47$ fm \cite{privcom}, and use the Woods-Saxon distribution $T_A(b)$ to calculate the saturation scale at other impact parameters:
\begin{equation}
Q^2_{s0}(b)=\frac{\bar{Q}_{s0}^2\ T_A(b)}{T_A(5.47\ \mbox{fm})}\ ,\quad\bar{Q}^2_{s0}=0.4\ \mbox{GeV}^2\ .
\end{equation}
Following the experimental analysis by the  STAR collaboration we set $|p_{1\perp}|>2$ GeV, $1\ \mbox{GeV}<|p_{2\perp}|<|p_{1\perp}|$ and $2.4< y_{1,2}< 4$, and require both hadrons to be neutral pions. Single-inclusive hadron production, used to normalize the coincidence probability, is calculated as in \cite{Albacete:2010bs}. We shall also calculate $CP(\Delta\phi)$ in p+p collisions in an analogous way, with the initial saturation scale $Q_0^2=0.2$ GeV$^2$ for the proton. Even though the applicability of our approach to the p+p case is questionable, the results obtained will be useful to interpret the phenomenon of monojet production observed in d+Au collision.

\begin{figure}
\includegraphics[width=8.1cm]{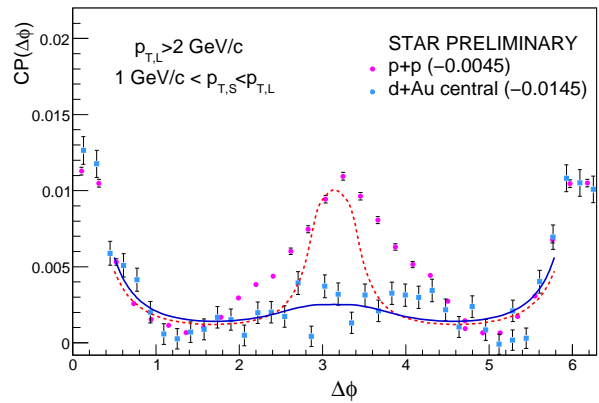}
\caption{The coincidence probability as a function of $\Delta\phi$ for central d+Au and p+p collisions. The preliminary data points by the STAR collaboration have been shifted vertically by constant amounts. Our results are displayed in solid lines.\vspace{-0.5cm}}\label{fig:star-comp}
\end{figure}

Our results for $CP(\Delta\phi)$ in central d+Au and in p+p collisions are displayed in Fig.~\ref{fig:star-comp}, along with preliminary data from the STAR collaboration \cite{Braidot:2010zh}. Several comments are in order. First, the disappearance of the away-side peak around $\Delta\phi\sim\pi$ in d+Au collisions exhibited by data is quantitatively well described by our CGC calculation. In turn our results for p+p collisions, where the applicability of the CGC formalism is not fully justified due to the smallness of the proton saturation scale, agree at a qualitative level with the presence of a well defined away-side peak in p+p collisions.

Analogous measurements of the coincidence probability in d+Au collisions at mid-rapidity, where $x_A$ is large, also display a clear back-to-back correlation \cite{Adams:2006uz,Adler:2006hi}. We thus conclude that {\it monojet} production is linked to the presence of a high gluon density in the target or, equivalently, to the fact that its saturation scale is comparable to the momenta of the produced hadrons. Similar conclusions have been obtained relying on a saturation model \cite{Tuchin:2009nf} (there, although different working assumptions are used, the presence of a saturation scale is also the crucial ingredient to successfully reproduce data). Our assumption of independent parton fragmentation in Eq.~\eqref{collfact} prevents us from extending our calculation to the near-side $\Delta\phi\sim0$ region. This would require the use of poorly constrained di-pion fragmentation functions. Moreover, the different measurements show little variation in the height and width of the near-side peak with varying colliding system or centrality, indicating that the near-side peak is not sensitive to saturation physics. Finally, since uncorrelated background has not been extracted from data, we have adjusted the overall normalization of data points by subtracting a constant shift.

\begin{figure}
\includegraphics[width=7.9cm]{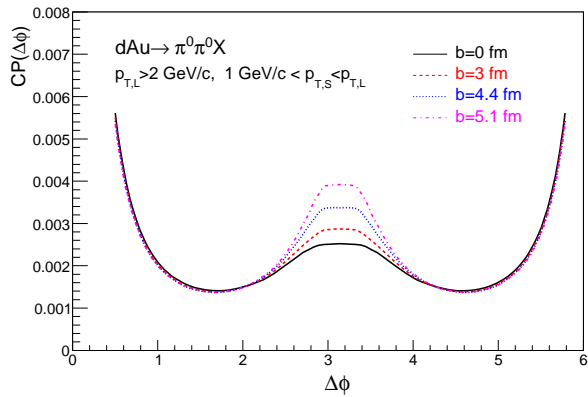}
\caption{The coincidence probability as a function of $\Delta\phi$ for different centralities of the d+Au collisions.\vspace{-0.5cm}}\label{fig:cent-dep}
\end{figure}

To further illustrate how the disappearance of the away-side peak is related to the increased gluon density, and quantify better the transition between p+p and central d+Au collisions, we show in
Fig.~\ref{fig:cent-dep} the centrality dependence of $CP(\Delta\phi)$. We predict that the near-side peak does not change with centrality, and that the away-side peak reappears for less central collisions. This is consistent with the fact that peripheral d+Au collisions are essentially  p+p collisions. We also predict that for higher hadron transverse momenta the away-side peak will reappear, as larger $x_A$ values will be probed
\cite{Marquet:2007vb}.

We conclude that the physics of {\it monojet} production is due to the interplay between the transverse momenta of the produced hadrons and the one acquired during the interaction with the nucleus. In the CGC approach presented here, the interaction with the nucleus is realized in a fully coherent way, and the momentum broadening is parametrically controlled by the $x$-dependent saturation scale of the nucleus. The latter, in turn, is described by means of the rcBK equation. While different interpretations of data may be possible, presently we are not aware of any alternative description of this phenomenon that does not invoke saturation effects. On the other hand, the simultaneous description of single and double inclusive hadron production at RHIC sets a constraining test to models. Here we obtain a good description of the d+Au data without modifying the few parameters already constrained by the analysis of single-inclusive hadron production presented in \cite{Albacete:2010bs}. Energy loss effects, which were a key ingredient in the alternative explanations \cite{alternatives} of the suppression of forward hadron yields in d+Au collisions, are not required in our approach in order to obtain a good description of forward di-hadron correlations. In this respect, the analysis presented here lends support to the idea that manifestations of the saturation regime of QCD have been observed at RHIC, and that such regime is well described by the CGC at its present degree of accuracy.

\begin{acknowledgments}

We are grateful to the STAR collaboration for allowing us to use their preliminary data. We would like to thank Les Bland and the STAR forward phyics group for many stimulating and useful discussions. The work of JLA is supported by a Marie Curie Intra-European Fellowship (FP7- PEOPLE-IEF-2008), contract No. 236376. CM is supported by the European Commission under the FP6 program, contract No. MOIF-CT-2006-039860.

\end{acknowledgments}

\end{document}